\documentclass[a4paper]{article}
\usepackage{graphicx}

\begin{document}

\begin{center}
{\bf \Large
Damage in  impact fragmentation}
\bigskip

{\large
N. Sator\footnote{sator@lptmc.jussieu.fr}$^{\star}$
and
H. Hietala$^{\ddag}$
}
\bigskip

\bigskip $^\star${ Laboratoire de Physique Th\'eorique de la Mati\`ere
  Condens\'ee, Universit\'e Marie et Pierre Curie (Paris 6), UMR CNRS
  7600 - 4 place Jussieu, 75252 Paris Cedex 05, France},

$^\ddag${ Department of Physics, University of Helsinki -
  P.O.Box 64, FI-00014 University of Helsinki, Finland}

\bigskip
\today
\end{center}

\begin{abstract}
  Using a simple and generic molecular dynamics model, we study the damage
  in a disc of interacting particles as the disc fragments upon impact
  with a wall. The damage, defined as the ratio of the number of bonds
  broken by the impact to the initial number of bonds, is found to
  increase logarithmically with the energy deposited in the
  system. This result implies a linear growth with damage for the
  total number of fragments and for the power law exponent of the
  fragment size distribution.
\end{abstract}



\bigskip

\section{Introduction}
\label{intro}

Because of its importance in natural phenomena and industrial
processes, fragmentation is of great interest. To a certain extent,
fragmentation phenomena depend on the particular features of the
object that is broken. However, some generic behaviours seem to be
shared by fragmenting systems whatever their size, material, or
typical interaction energy. For instance, as observed in a large
variety of experiments
\cite{turcotte86,matsui82,meibom96,oddershede93,kadono97,kadono02,campi00,astrom04,wittel04,moukarzel07}
and natural phenomena \cite{turcotte86,kaminski98,oddershede98}, the
fragment size distribution frequently exhibits a power law behaviour,
the origin of which is still unknown.

Despite its simple and rather everyday aspect, fragmentation is a
complex and distinctly far-from-equilibrium process which makes it
difficult to understand theoretically.  Attempts have been made to
derive a power law distribution using analytical models, such as
sequential fragmentation \cite{astrom04} and energy-balance theory
\cite{grady08}. However, these approaches do not succeed in explaining
the variety of values of the power law exponent observed in
experiments.  On the other hand, fragmentation is beneficially studied
with computer simulations
\cite{holian88,ching99,diehl00,astrom00,thornton96,mishra01,kun99,behera05,wittel08,myagkov05,sator08}.
Indeed, this approach can grasp the complexity and the dynamical
properties of fragmentation by taking into account the various
parameters that may influence the process --- and consider even them
one by one.

The main goal of our studies is to highlight and understand the
behaviours observed in a wide range of different fragmenting
systems. To encompass this variety, we recently proposed an elementary
molecular dynamics (MD) model \cite{sator08} which, due to its
simplicity, provides a very generic frame of reference for
fragmentation studies.  Therefore, this model is not representative of
any specific material. We applied it to investigate the fragmentation
of a two-dimensional disc of interacting particles upon impact with a
wall \cite{sator08}. One of the main results of this study is, that
the power law exponent of the fragment size distribution increases
logarithmically with the energy deposited in the system. This
behaviour is in agreement with experimental results
\cite{matsui82,moukarzel07} and we expect it to be generic for
fragmentation phenomena.

In the present paper, we extend this study by addressing the issue of
damage occurring in a fragmenting disc. In particular, we will show
how the breaking of bonds at the microscopic level is related to
macroscopic quantities such as the fragment size
distribution. Dynamical quantities dealing with the spreading of
damage during the fragmentation process will also be discussed.

The paper is organized as follows: in section 2, we describe the model
and the simulation procedure ; in section 3 we review the main results
concerning the fragment size distribution ; section 4 is devoted to the
study of damage ; in section 5 we discuss the fragmentation energy. We
conclude in section 6.

\section{Model and simulations}
\label{sec:model}

In the present study, the fragmenting system is a disc made up of
$N=1345$ particles placed on a two-dimensional triangular lattice.  In
the previous article \cite{sator08}, we showed by a finite size
scaling approach that the fragmentation features of the system are not
sensitive to the number of particles.  The cohesion of the system is
ensured by a central two-body Lennard-Jones type of potential
\begin{equation} 
\label{eq.1} 
\mathrm{v(r_{ij})}=v_0 \epsilon
\bigg[\bigg(\frac{\sigma}{\mathrm{r_{ij}}}\bigg)^{a}-\bigg(\frac{\sigma}{\mathrm{r_{ij}}}\bigg)^{b}\bigg].
\end{equation}
Here $\mathrm{r_{ij}}$ is the distance between particles $i$ and $j$
and the two constants, $\epsilon$ and $\sigma$, are the depth of the
potential well and the diameter of the particles, respectively. The
three parameters, $v_0=107.37$, $a=80$ and $b=78$, were chosen to
obtain a strong repulsion at contact and a very short range of
attraction (around $0.1\,\sigma$ in addition to the particle
diameter). These are the features that one expects to be dominant
within a brittle solid at mesoscopic scales. Note that the behaviour
of the system under fragmentation is not sensitive to a particular
choice of the parameters $v_0$, $a$, and $b$, that is, to the range of
attraction \cite{sator08}.

The fragmentation process was studied by means of 2D molecular
dynamics simulations utilizing the Verlet
algorithm~\cite{verlet67,frenkel01} at constant energy. The time step
is chosen as $\delta t=0.0001 \, t_0$, ensuring the conservation
  of total energy up to 0.001 $\%$. Here $t_0=\sqrt{\epsilon
    \sigma^2/m}$ is the unit of time and $m=1$ the particle mass.  In
  the following, times, lengths, velocities, and energies are
  expressed in the units of $t_0$, $\sigma$, $\sigma/t_0$, and
  $\epsilon$ respectively.

The disc of diameter $L=40 \sigma$ is constructed, rotated by a random
angle and launched into a solid wall with a given impact velocity $\bf
{V}$.  The particles of the disc interact with the wall through the
repulsive part of the potential $v(r_{ij})$ (for details, see
\cite{sator08}). For each value of the impact velocity, we performed
1000 runs while sampling the initial angle of rotation
uniformly. Simulations were stopped at $t=100$ (i.e. after $10^6$ MD
time steps), when the fragmentation process had already reached a
steady state \cite{sator08}.

During the simulation, fragments are defined as self-bound clusters of
particles~\cite{sator03}. In other words, two particles are bonded if
their relative kinetic energy is lower than the absolute value of
their interaction energy.  Once the distance $r_{ij}$ between two
particles $i$ and $j$ becomes larger than the attractive range ($ \sim
1.2\sigma$) of the potential, their potential energy is set from
$\epsilon$ to $0.001 \epsilon$. This value reduces the interaction
between these two particles to an almost repulsive one and thus the
possible recombination is prevented.

In this article, we measure the energy deposited in the system with a
dimensionless control parameter $\eta$, defined as the ratio of the
initial kinetic energy to the potential energy of the disc:
$$
\displaystyle \eta=\frac{mV^2}{2 \; |e_{pot0}|}.
$$
Here $e_{pot0} =-2.87 \epsilon$ is the initial potential energy per
particle in the disc. When the kinetic energy is large enough compared
to the cohesion of the disc, fragmentation occurs.

The same numerical experiment --- a disc of interacting particles
fragmenting upon impact with a wall --- has been previously
investigated by means of more complex models.  For instance, Thornton
and his collaborators \cite{thornton96} used primary particles
interacting through autoadhesive and frictional forces that act as
functions of contact area. The particles, set initially in random
locations, are pulled together by applying a centripetal force. Hence
in their model, the fragmenting agglomerate has an irregular shape and
contains inherent structural defects.  Likewise, in the model proposed
by Kun, Herrmann, and their collaborators
\cite{kun99,behera05,wittel08}, the primary particles are rigid
randomly shaped convex polygons. These polygons are connected with
beams that take into account elastic, shear, and torque interactions.

The asset of the present study is the use of a minimal fragmentation
model since it provides a generic frame of reference to which one can
compare the effects of different parameters.  Accordingly, the system
considered in this article is perfectly ordered and homogeneous, but
disorder and heterogeneities could be easily taken into account to
investigate their effects.

\section{Fragment size distribution }
\label{sec:distribution}

\begin{figure}[h]
 \includegraphics[angle=0,scale=.4]{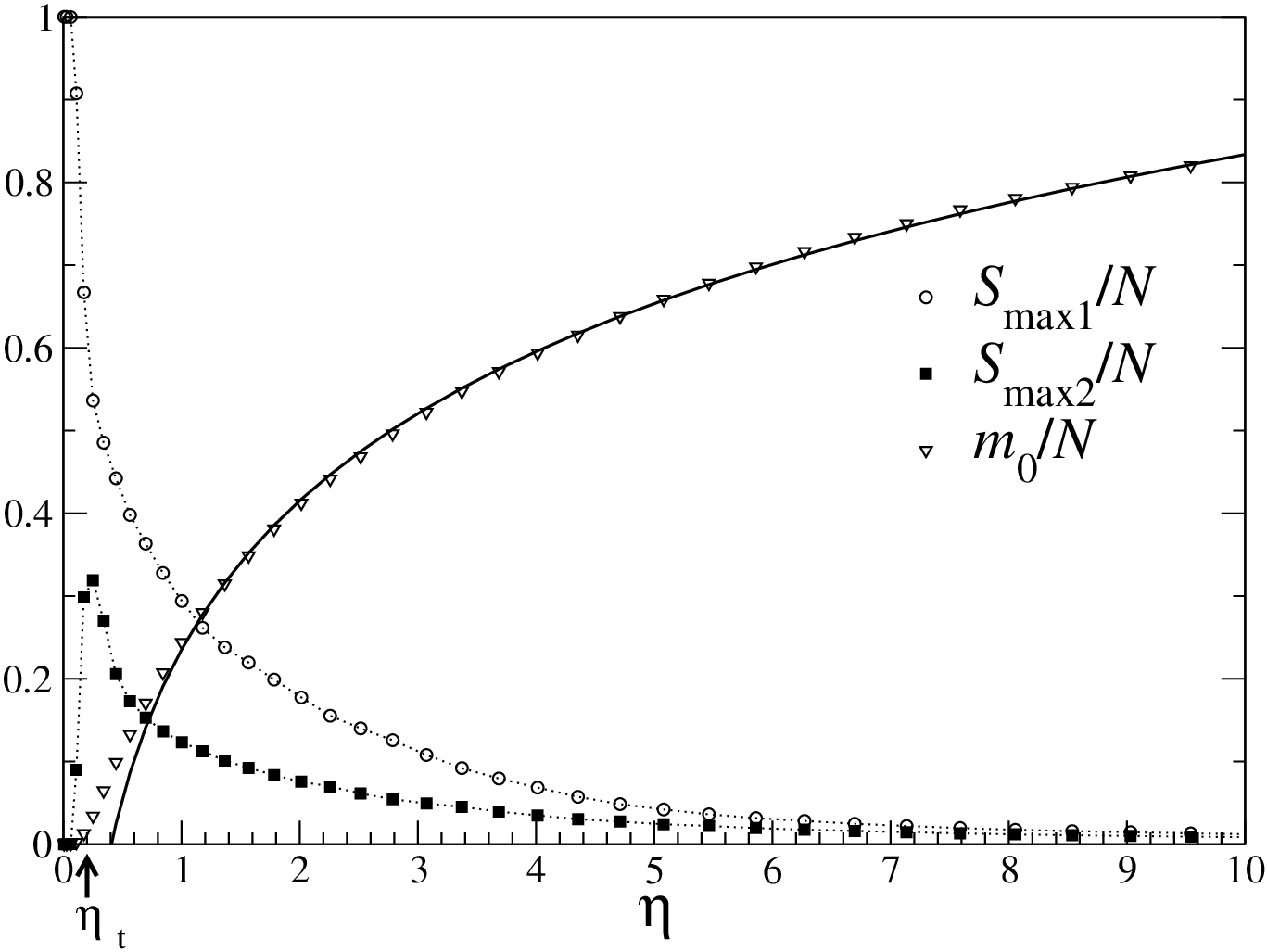}
 \caption{\it The mean size of the first and second largest fragments,
   $S_{max1}/N$ and $S_{max2}/N$, and the total number of fragments
   $m_0 /N$ as functions of $\eta$ at time $t=100$. The full line
   corresponds to a logarithmic fit (see eq. (\ref{eq:m0})). Dotted lines are a guide for the
   eyes. The threshold energy $\eta_t$ is indicated by an arrow.}
\label{fig:1}       
\end{figure}

For a given amount of energy deposited in the system, we calculated
the fragment size distribution --- the number $n(s)$ of fragments made
up of $s$ particles --- averaged over 1000 runs. In addition to this,
we calculated the normalized mean size of the largest and the second
largest fragment --- $S_{max1}$ and $S_{max2}$ --- as well as the
total number of fragments $m_0$. These last quantities are plotted in
Fig. \ref{fig:1} as functions of $\eta$ at the end of the simulation
($t=100$).  As we found in the previous study \cite{sator08},
$S_{max1}$ decreases with $\eta$, whereas $S_{max2}$ is peaked at
$\eta_t =0.19$ ($V_t=1.05$). When the amount of energy deposited in
the system is larger than this threshold value $\eta_t$, the fragment
size distribution exhibits a power law behaviour with an exponent
$\tau$: $\displaystyle n(s) \sim s^{-\tau}$.

\begin{figure}[h]
 \includegraphics[angle=0,scale=.42]{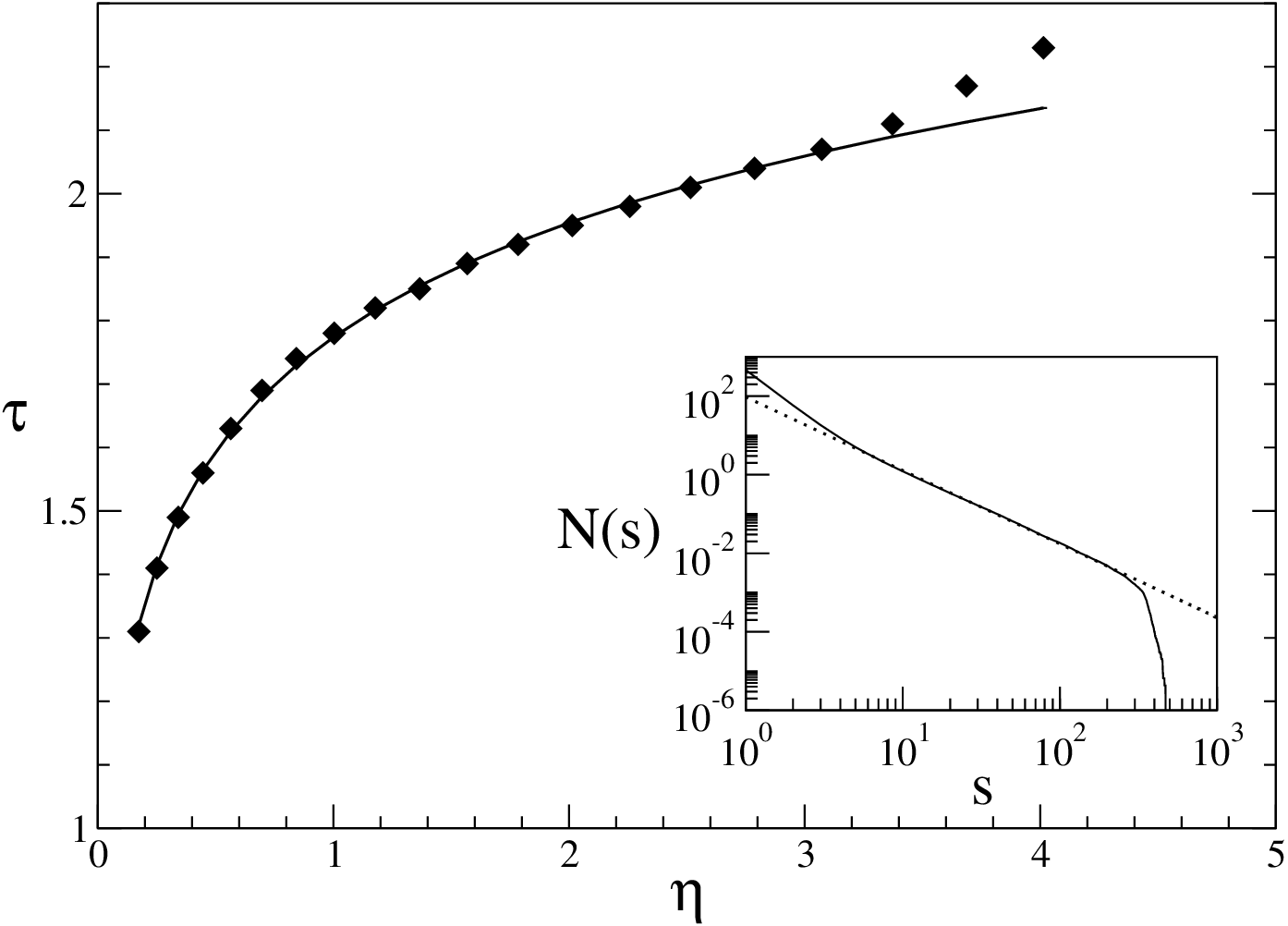}
 \caption{\it Power law exponent $\tau$ as a function of $\eta$. The full
   line is a logarithmic fit (see eq. (\ref{eq:tau})).  Cumulative
   fragment size distribution for $V=3$ ($\eta=1.57$) at $t=100$ is plotted in the
   inset (the dotted line is a power law function with $\tau =1.87$).}
\label{fig:2}       
\end{figure}

To estimate the exponent $\tau$, we employed statistical procedures
proposed by Clauset \textit{et al.}  \cite{clauset07}, which are based
on the method of maximum likelihood. Note that, as $\eta$ increases to
values much larger than $\eta_t$, the power law region of $n(s)$
narrows due to the finite size of the fragmenting disc, and the
determination of $\tau$ becomes difficult. Nevertheless, the power law
fit is quite good for $\eta_t \le \eta \le 4$, (see the fragment size
distribution plotted in the inset  of Fig. \ref{fig:2} ).

Furthermore, as illustrated in Fig. \ref{fig:2}, the exponent $\tau$
of the power law exhibits a logarithmic increase with the energy
deposited in the system:
\begin{equation}
\label{eq:tau}
\tau = \alpha_{\tau} \ln \frac{\eta}{\eta_t}+\beta_{\tau},
\end{equation}
where $\alpha_{\tau}=0.26 \pm 0.01$ and $\beta_{\tau}=1.34 \pm
0.01$. When $\tau$ is given as a function of the impact velocity, the
coefficients $\alpha_{\tau}$ and $\beta_{\tau}$ are slightly different
than those obtained in the previous article \cite{sator08}. This is
due to the fact that, in the present paper, we use a more rigorous
method \cite{clauset07} for fitting.

Similar to the size of the largest fragment, the total number of
fragments $m_0=\sum_s n(s)$ is a measure of the degree of break-up in
the fragmentation process. Figure \ref{fig:1} displays also $m_0$ as a
function of $\eta$. As expected, the total number of fragments
increases with the initial kinetic energy.  Moreover, $m_0$ presents a
linear behaviour with $\eta$ for $\eta \le 1$, and a logarithmic
behaviour for $1 \le \eta \le 10 $, that is, up to the highest impact
energies considered in this work:
\begin{equation}
\label{eq:m0}
\frac{m_0}{N}= \alpha_{m_0} \ln \eta +\beta_{m_0}.
\end{equation}
Here $\alpha_{m_0}=0.26 \pm 0.01$, i.e., the same coefficient as for
$\tau$, and $\beta_{m_0}=0.23 \pm 0.01$.  Note that Behera {\it et
  al.}  \cite{behera05} found the same linear and logarithmic
behaviours while studying the impact fragmentation of a disc with a
more sophisticated model, that takes into account also the elastic,
shear and torque interactions between particles.

\section{Damage}
\label{sec:damage}

\begin{figure}[h] 
\includegraphics[angle=0,scale=.4]{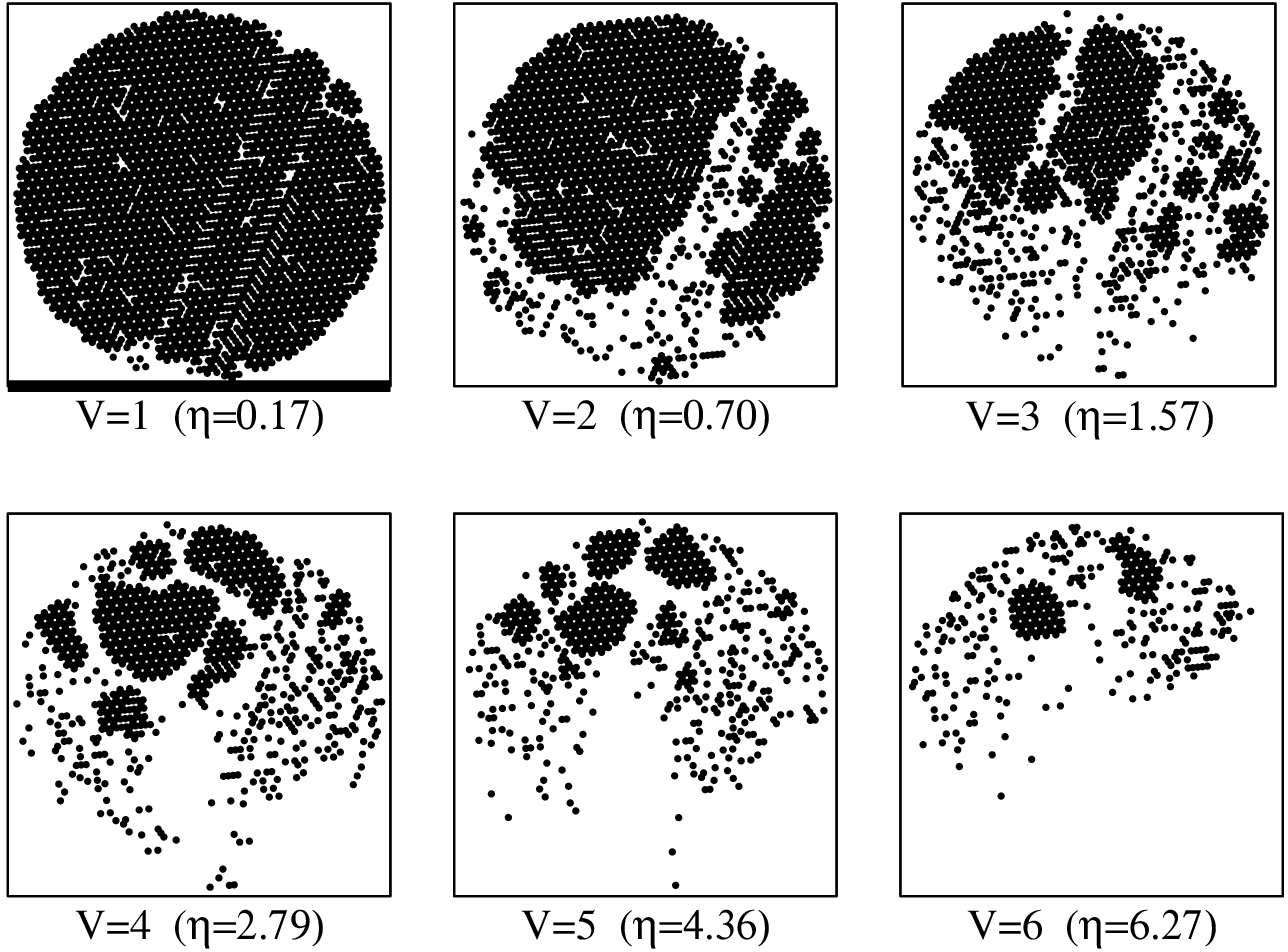}
\caption{\it Snapshots at the end of the fragmentation process ($t=100$)
  for various impact velocities and the same initial rotation
  angle. Each dot represents a bond between two particles localized at
  its initial position before the impact. The horizontal wall is
  depicted at the bottom of the top left figure.}
\label{fig:3}       
\end{figure}

The fragment size distribution discussed in the previous section
reflects the degree of damage caused by the impact. Furthermore, the
increase of the total number of fragments with the impact velocity
would be, and is, a direct and intuitive representation of increasing
damage. However, both of these are macroscopic quantities. The
microscopic level of the process is of greater interest if we want to
understand how the fragmentation occurs. In our model, cracks and
fragments are formed by the breaking of bonds between neighboring
particles. This is illustrated in the snapshots of Fig. \ref{fig:3},
where the bonds that still exist at the end of the fragmentation
process ($t=100$) are shown in their initial locations before the
impact. Fragments can be discerned as clusters of the remaining bonds
(except the monomers, of course).

At low initial kinetic energy ($V=1$, $\eta =0.17$), the disc suffers
mainly from internal damage. Some internal bonds are broken and a few
fragments are formed, mainly monomers in the impact zone (bottom of
the figure). Oblique lines of broken bonds starting from the impact
zone can be seen. As the energy deposited in the system increases
($V=2$, $\eta =0.70$), oblique cracks propagate through the system
from the impact zone to the edge of the disc and fragmentation in the
proper sense of the word occurs. The largest fragment is localized in
the top part of the disc.  Most of the energy is dissipated in the
impact zone, producing quantities of monomers.  At higher energies ($V
\ge 3$ , $\eta \ge 1.57$), cracks perpendicular to the former oblique
ones form and create smaller fragments by merging. As energy increases
further, fragments become smaller but are of various sizes, in
qualitative agreement with a power law fragment size distribution. It
is interesting to note, that the shape of the fragments is irregular
and rough without memory of the underlying triangular lattice
structure of the disc. To conclude, the crack patterns shown in
Fig. \ref{fig:3} are consistent with the ones obtained with more
complex models \cite{thornton96,behera05}.

\begin{figure}[h] 
\includegraphics[angle=0,scale=.4]{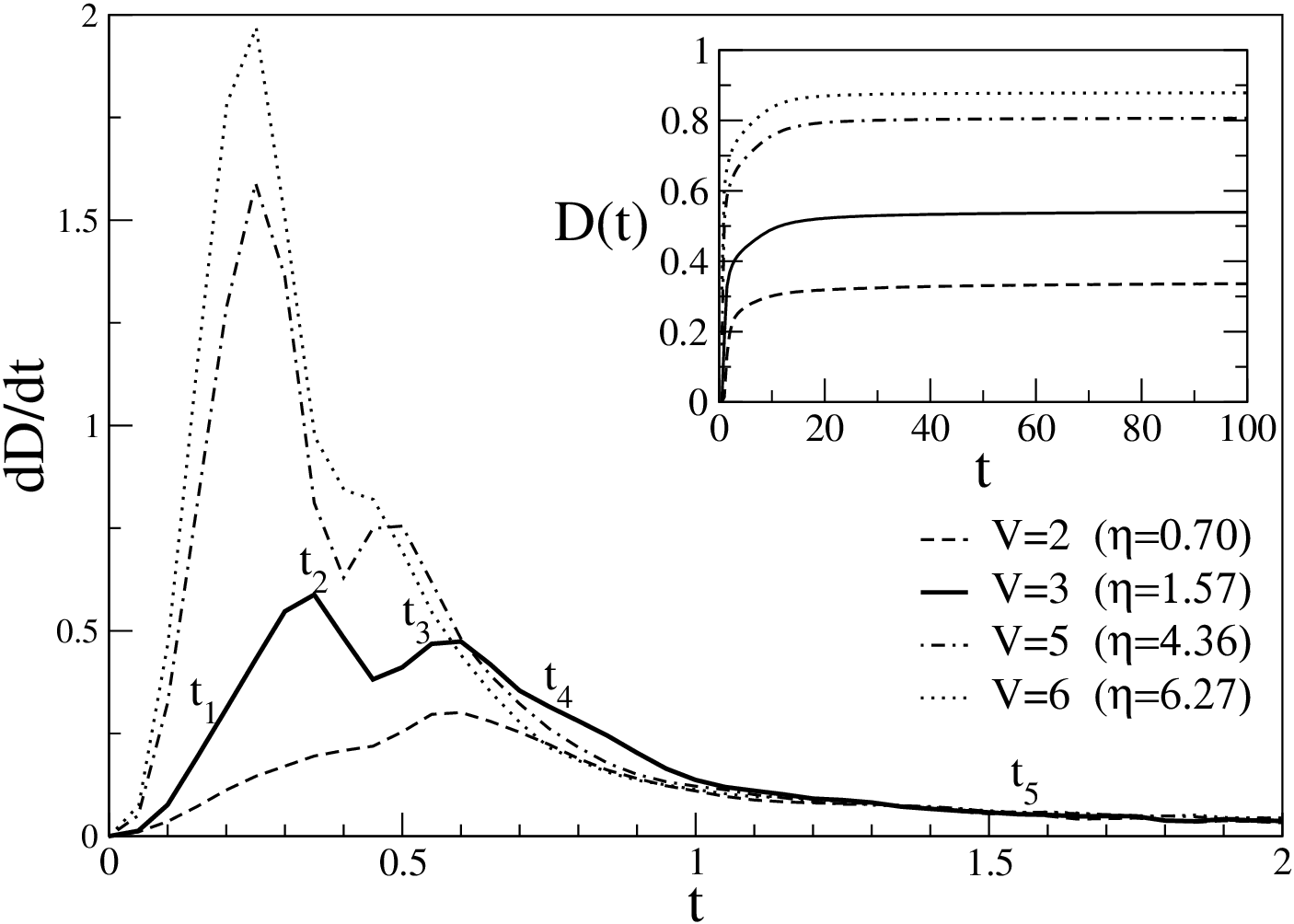}
\caption{\it Damage rate ($\frac{d D}{ dt}$) as a function of time for
  various impact velocities. The times $t_i$, for $i=1,2,3,4,5$, along
  the curve for $V=3$ correspond to the snapshots of
  Fig. \ref{fig:5}. In the inset, $D(t)$ as a function of time for the
  same impact velocities.}
\label{fig:4}       
\end{figure}

To quantify the damage produced during the impact fragmentation,
Thornton and his collaborators \cite{thornton96} defined the damage
ratio $D(t)$ as the average ratio of the number of broken bonds at a
given time $t$, to the total number of initial bonds in the disc.
This definition of damage is quite natural at a microscopic scale and
can be easily applied to many kinds of fragmenting systems. In the
inset of Fig. \ref{fig:4} we plot the damage ratio for various impact
velocities. At the moment of impact ($t=0$), $D(t)$ first increases
dramatically, then at a slower rate, and eventually tends to its final
value $D_f$, showing that the fragmentation process reaches a steady
state. In fact, at later times ($t>50$), only few particles evaporate
from exited fragments, increasing very slightly the damage ratio.

In order to better understand the evolution of damage, we calculated
the damage rate, $\displaystyle \frac{d
  D(t)}{dt}$.  As shown in Fig \ref{fig:4}, the damage rate increases
drastically, reaches a maximum, and then decreases slowly to zero.

\begin{figure}[h]
\includegraphics[angle=0,scale=.42]{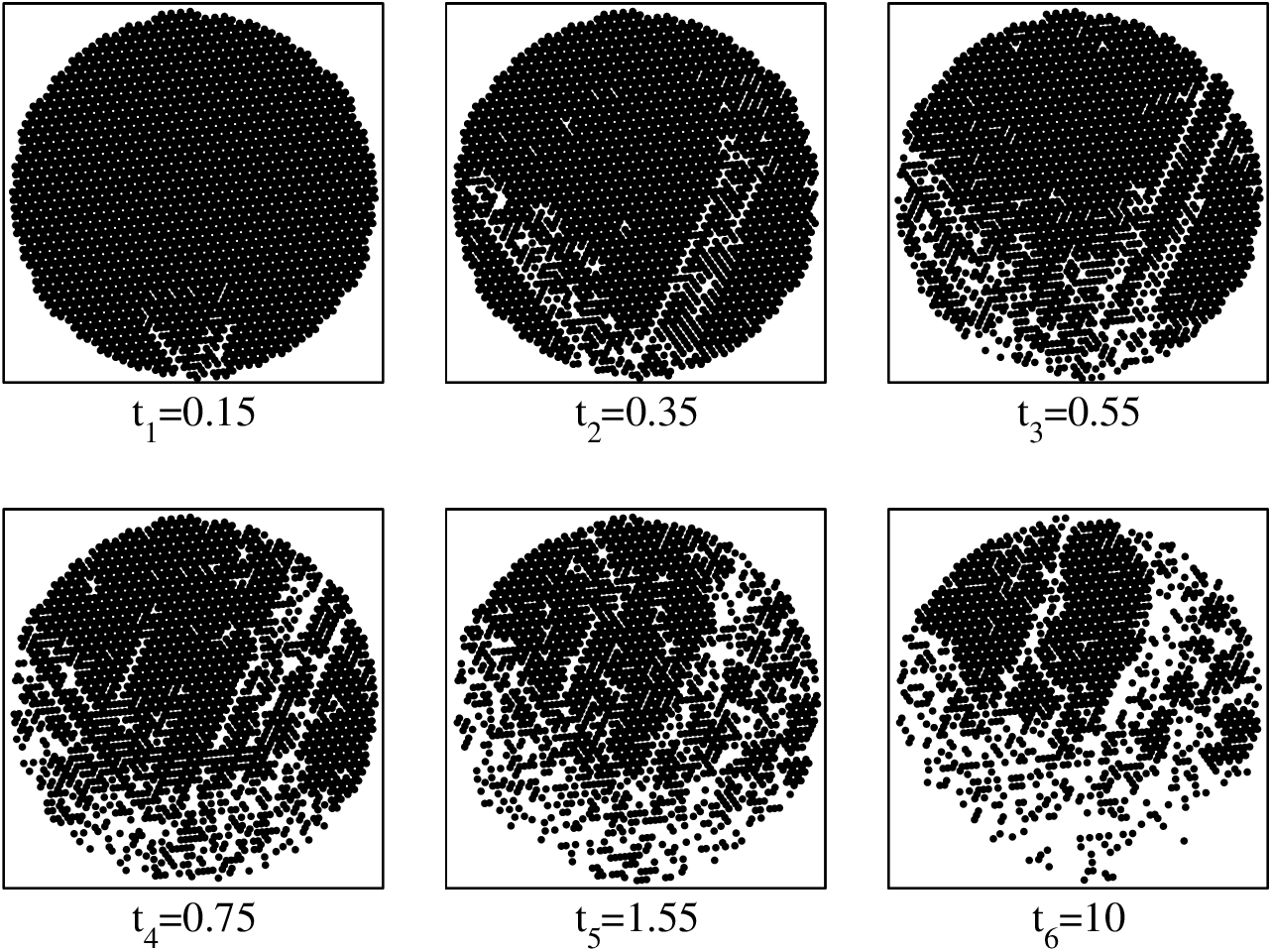}
\caption{\it Snapshots at various times for $V=3$ ($\eta =1.57$). Each dot
  represents a bond between two particles placed at its initial
  position before the impact.}
\label{fig:5}       
\end{figure}

This evolution is further illustrated by the snapshots in Fig
\ref{fig:5} of the fragmenting disc for the impact velocity $V=3$
($\eta=1.57 $).  Just after the impact, at $t_1=0.15$, two oblique
cracks start to propagate from the impact zone. At $t_2=0.35$, damage
increases in the impact zone and the oblique cracks become wider and
continue their advance through the disc. As a consequence the damage
rate increases swiftly. When the two cracks have reached the edge of
the disc, the damage rate is maximal.  Then, cracks perpendicular to
the oblique ones are formed ($t_3=0.55$). Damage spreads into the disc
and damage rate decreases ($t \ge t_3$). The local minimum of the
damage rate ($t_2 \le t \le t_3$) is due to a sudden increase of the
kinetic energy per particle (not shown) when the disc rebounds, after
being compressed (this is observed also for higher impact velocities,
see the curve for $V=5$). Finally, bond breakings create fragments
whose surfaces evolve by evaporating particles ($t \ge t_5 =1.55$). As
can be seen by comparing Fig. \ref{fig:5} at $t_6=10$ and
Fig. \ref{fig:3} at $t=100$, fragments already have their overall
shape at $t=10$. Note that the lattice structure influences the first
oblique crack propagation. However, these linear cracks are not
sufficient to dissipate the impact energy and secondary winding cracks
form within the disc without revealing the underlying lattice
structure.

As illustrated by the curves in Fig. \ref{fig:4}, as the impact
velocity increases, the damage rate increases faster and reaches a
maximal value when the oblique cracks have propagated through the
system. The position of the peak depends only slightly on the energy
deposited in the system, while it is mostly determined by the
direction of propagation of the cracks. Indeed, as the speed of sound
in the disc is estimated to be around 100 \cite{sator08}, a crack of
length $40 \sigma$ (the diameter of the disc) reaches the edge at time
$t \simeq 0.4$. This is in agreement with the position of the peaks in
Fig. \ref{fig:4}.  Note that the regime of evaporation is reached at
the same time, around $t\simeq 1$, regardless of the energy deposited
in the system.

\begin{figure}[h]
 \includegraphics[angle=0,scale=.4]{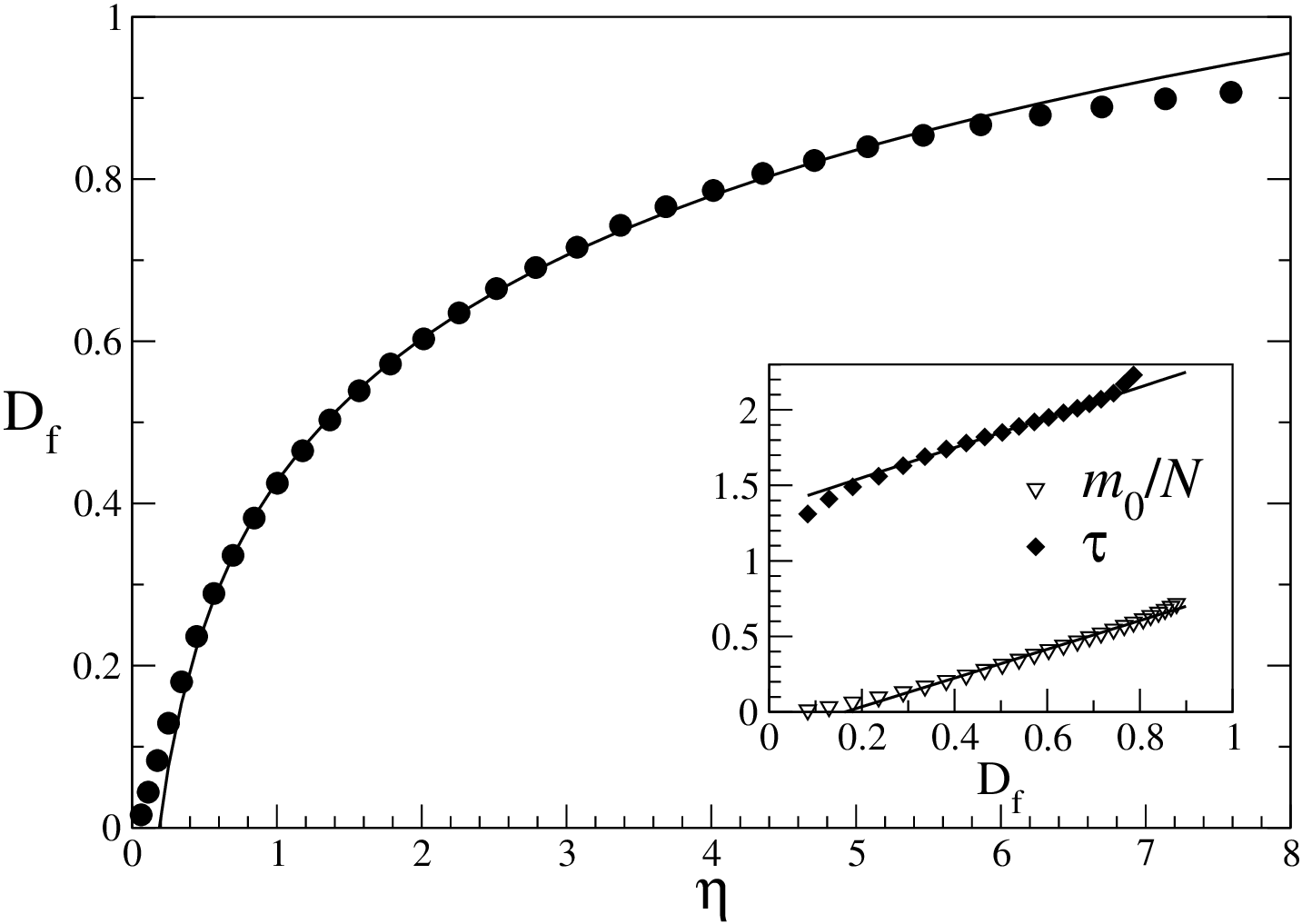}
 \caption{\it Final damage ratio $D_f$ at $t=100$ as a function of
   $\eta$. The full line is a logarithmic fit (see eq. (\ref{eq:dam})).
   The power law exponent $\tau$ and the total number of fragments
   $m_0 /N$ are plotted in the inset as functions of damage ratio
   $D_f$ (the full lines are linear fits).}
\label{fig:6}       
\end{figure}

Having studied the evolution of damage, we discuss how
the final damage ratio $D_f$ at the end of the process behaves as a function of the
initial kinetic energy.  As illustrated by the curve in
Fig. \ref{fig:6}, the final damage ratio $D_f$ is very well fitted by
a logarithmic function for $0.4 \le \eta \le 6$, or equivalently for
$0.2 \le D_f \le 0.9$:
\begin{equation}
\label{eq:dam}
D_f= \alpha_{D_f} \ln \frac{\eta}{\eta_t}.
\end{equation}
Here $\alpha_{D_f }=0.256 \pm 0.005$, i.e., almost the same
coefficient as for $\tau$ and for $m_0/N$. For higher energies ($\eta
\ge 6$), $D_f$ grows more slowly than the logarithmic function,
reflecting a lower efficiency of the fragmentation process. Therefore,
even at very high initial kinetic energies, some small fragments still
remain. For instance, at $V=8$ ($\eta=11.15$), $S_{max1} \simeq 13$.

The same logarithmic behaviour of $D_f$ was reported by Thornton {\it et al.}
\cite{thornton96} and Behera {\it et al.} \cite{behera05}. For the
former model, they found $\alpha_{D_f} \simeq 0.1$, depending on the surface energy of
the disc. For the latter one, $\alpha_{D_f} \simeq 0.15$, as can be
estimated from Fig. 8 of reference \cite{behera05}.

According to equations (\ref{eq:tau}), (\ref{eq:m0}), and
(\ref{eq:dam}), the power law exponent $\tau$, the total number of
fragments $m_0 /N$, and the damage ratio $D_f$ all exhibit a
logarithmic dependence on $\eta$ with the same coefficient $\alpha$,
in a certain range of energy. Consequently, we expect a linear
relation between $\tau$ and $D_f$, and between $m_0/N$ and $D_f$. This
is plotted in the inset of Fig. \ref{fig:6}.  A linear fit gives
\begin{equation}
\label{eq:taudam}
\tau = D_f+1.35,
\end{equation}
for $0.2 \le D_f \le 0.7$ and 
\begin{equation}
\label{eq:m0dam}
m_0/N = 0.96D_f-0.16,
\end{equation}
for $0.2 \le D_f \le 0.9$, with a slope slightly less than 1. Note
that using the data presented by Behera {\it et al.} \cite{behera05},
we found a linear relation between $m_0$ and $D_f$ as well.

\section{Fragmentation energy}
\label{sec:energy}

A large part of the total energy consumed in industry is used to
achieve size reduction of materials. To estimate the efficiency of the
fragmentation process studied in this article, we calculated the
\emph{fragmentation energy}.  Noting that no frictional interaction is
taken into account in the present model, it follows from the
conservation of energy that
 \begin{equation}
\label{eq:nrj}
e_{pot0}+\frac{1}{2}mV^2  =e_{pot} +e_{kin},
\end{equation}
where $e_{pot}$ and $e_{kin}$ are the average potential and kinetic
energies per particle at the end of the fragmentation process, and
$e_{pot0} =-2.87 \epsilon$ is the initial potential energy per
particle.  In other words, the initial kinetic energy is used to break
bonds and create fragments (potential energy) and to move the
fragments apart (kinetic energy). Consequently, we define the
fragmentation energy $e_{frag}$ as the potential energy needed to form
the fragments: $e_{frag}=e_{pot} -e_{pot0}$.

It is easy to see that the fragmentation energy is related to the
final damage ratio $D_f$. Using a linear fit, we found for $D_f \ge
0.1$:
\begin{equation}
\label{eq:nrj}
e_{frag}=2.75 D_f +0.12.
\end{equation}
Due to surface effects, the fragmentation energy is lower than
$-e_{pot0}D_f$ when the disc is broken into fragments. When the disc
is completely shattered and $D_f=1$, we have $e_{frag} =-e_{pot0}$, as
expected.

\begin{figure}[h] 
\includegraphics[angle=0,scale=.4]{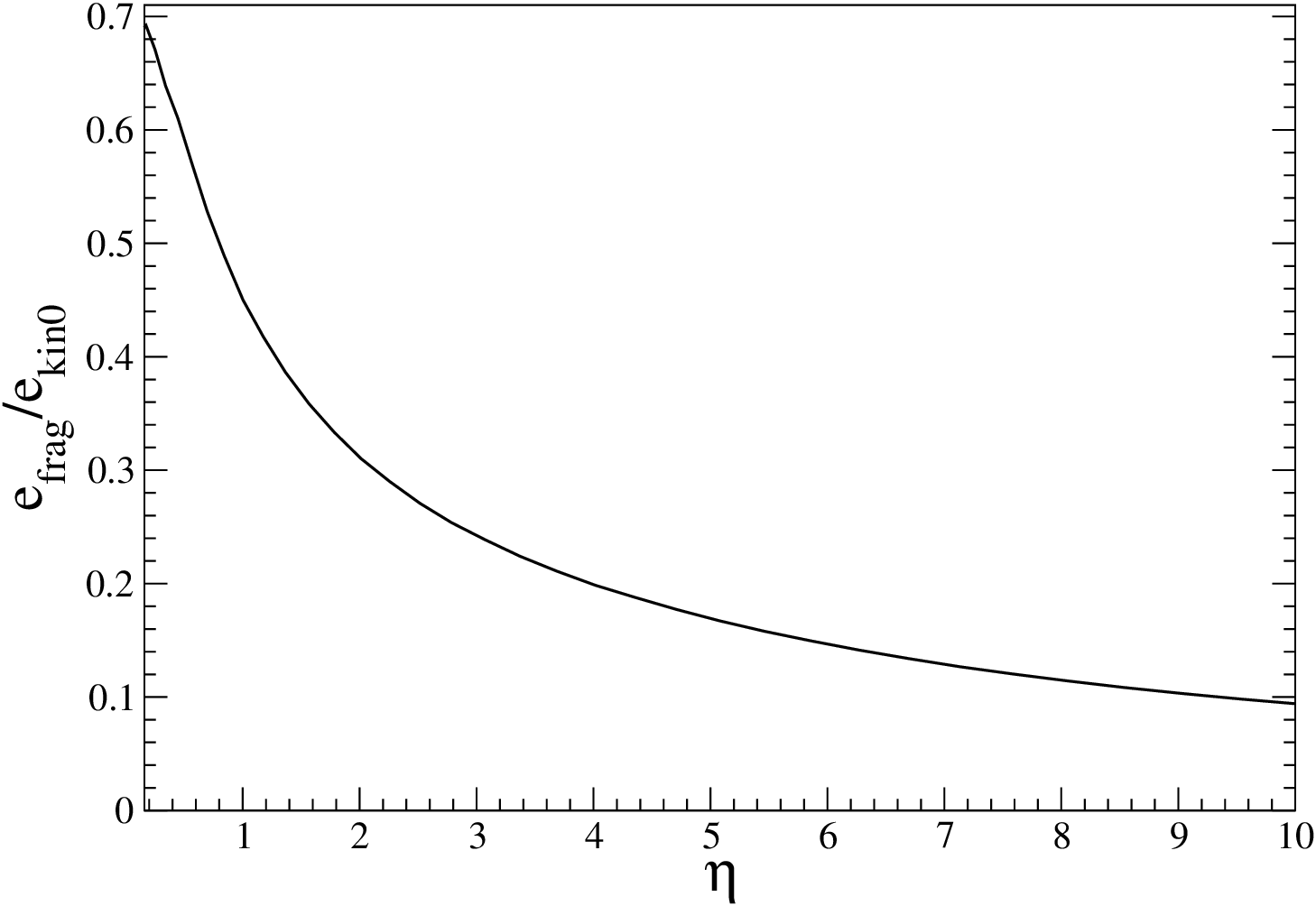}
\caption{\it Fragmentation energy divided by the initial kinetic energy,
  $e_{frag}/e_{kin0}$, as a function of $\eta$, for $\eta \ge \eta_t$.}
\label{fig:7}       
\end{figure}

The \emph{efficiency} of the fragmentation process can be estimated by
calculating the fraction of energy used to create the fragments, that
is, $e_{frag}/e_{kin0}$ where $e_{kin0}=\frac{1}{2}mV^2$ is the
initial kinetic energy per particle. This quantity is plotted in
Fig. \ref{fig:7} as a function of $\eta$ in the fragmentation regime,
i.e., for $\eta > \eta_t$. The percentage of the initial kinetic
energy that is used to form the fragments decreases from $70 \%$ to
$10 \%$ as $\eta$ increases from $\eta_t$ to 10. At high impact
velocities, most of the energy is then spent on moving the fragments
apart.  If friction would be taken into account, the efficiency of the
fragmentation process would be even lower.

\section{Conclusion}
\label{sec:conclusion}

The model we propose is generic in the sense that it contains the
essential physical features for investigating fragmentation phenomena.
Its simplicity rests on a basic cohesive interaction between circular
particles in two dimensions. Furthermore, molecular dynamics
calculations allow us to determine the physical quantities involved
and moreover, to study their evolution during the fragmentation
process.

In this article, we have investigated the damage inflicted on a disc
of interacting particles as the disc fragments upon impact with a wall
--- the damage being defined as the percentage of broken bonds between
particles. Its propagation into the disc evidently depends on the
energy deposited in the system, but the damage rate was found to be
independent of the impact energy shortly after the
collision. Furthermore, the percentage of energy actually used to
fragment the system decreases with the energy deposited in the system,
reflecting the low efficiency of the fragmentation process at high
impact energy.

In particular, we have shown that the power law exponent $\tau$ of the
fragment size distribution, the total number of fragments, and the
damage share the same logarithmic behaviour as functions of the energy
deposited in the system. Damage cannot be directly measured by
experiments, but the logarithmic behaviour of $\tau$ is observed in
the fragmentation of rocks \cite{matsui82} and liquid droplets
\cite{moukarzel07}. Surprisingly, these two very different systems
exhibit the same fragmentation behaviour. This suggests that at high
impact energies, the crack propagation and fragmentation phenomena may
not be sensitive to the particular structure of the system.  As an
interesting consequence of this same logarithmic behaviour, the power
law exponent and the total number of fragments are proportional to the
damage. These simple relations, that we expect to be generic, may
guide us to a better understanding of the underlying mechanisms of
fragmentation phenomena.

\end{document}